\documentclass[a4paper,11pt]{article}

\pdfoutput=1 

\usepackage{jinstpub} 
\usepackage{color}
\usepackage{lineno}

\title{Response of the GAGG(Ce) scintillator to charged particles
compared with the CsI(Tl) scintillator}

\author[a,1]{T.~Furuno,\note{Corresponding author.}}
\author[b]{A.~Koshikawa,}
\author[a]{T.~Kawabata,}
\author[c]{M.~Itoh,}
\author[d,e]{S.~Kurosawa,}
\author[f]{T.~Morimoto,}
\author[g]{M.~Murata,}
\author[a]{K.~Sakanashi,}
\author[f]{M.~Tsumura,}
\author[d,e]{and A. Yamaji}

\affiliation[a]{Department of Physics, Osaka University, 
  Toyonaka, Osaka 560-0043, Japan}

\affiliation[b]{Graduate School of Information Sciences, 
        Tohoku University, Sendai, Miyagi 980-8578, Japan}

\affiliation[c]{Cyclotron and Radioisotope Center, Tohoku University, 
  Sendai, Miyagi 980-8578, Japan}

\affiliation[d]{Institute for Materials Research, Tohoku University, 
  Sendai, Miyagi 980-8577, Japan}

\affiliation[e]{New Industry Creation Hatchery Center, Tohoku University, 
  Sendai, Miyagi 980-8579, Japan}

\affiliation[f]{Department of Physics, Kyoto University,
  Sakyo, Kyoto 606-8502, Japan}

\affiliation[g]{Research Center for Nuclear Physics, Osaka University,
  Ibaraki, Osaka 567-0047, Japan}

\emailAdd{furuno@ne.phys.sci.osaka-u.ac.jp}

\abstract{
  GAGG(Ce) is a novel scintillator with a fast response
  and high light output without a hygroscopic nature.
  It is expected to be a useful detector for charged particles
  at high-counting rates.
  However, the response of the GAGG(Ce) scintillator
  to charged particles has not been fully examined.
  In the present work, the light output and energy resolution
  of the GAGG(Ce) scintillator were measured for protons and alpha particles at
  $E_{p}=5$--68 MeV and $E_{\alpha}=8$--54 MeV as well as
  gamma rays at $E_{\gamma}=662$ keV from a $^{137}$Cs source.
  The results were compared with those of the CsI(Tl) scintillator.
  The scintillation efficiencies $dL/dE$ of the GAGG(Ce) and CsI(Tl)
  scintillators were obtained and parametrized
  as a function of linear energy transfer $dE/dx$.
}


\begin{document}
\maketitle
\flushbottom

\section{Introduction}
\label{sec:intro}

Ce-doped Gd$_{3}$Al$_{2}$Ga$_{3}$O$_{12}$
[GAGG(Ce)] is a novel scintillation material
developed by Institute for Materials Research, Tohoku University
and Furukawa CO., LTD \cite{Kamada2011, KAMADA201288, Kurosawa2014}.
The main decay time constant of the GAGG(Ce) scintillator is
reported as 60--100 ns \cite{Iwanowska2013,Sibczynski2015,Gundacker2018},
thus it seems to be appropriate for measurements at high-counting rates.
The GAGG(Ce) scintillator has been employed as gamma-ray detectors
for numerous purposes such as
Compton camera \cite{Yamamoto2016, Hosokoshi2019},
positron emission tomography \cite{gagg_pet}, and
radiation imaging \cite{Kamada2014}.

So far, the Tl-doped alkali iodides such as NaI(Tl) and CsI(Tl) scintillators
have been commonly used
not only for gamma-ray measurements \cite{Takeuchi2014,Cardella2015}
but also for charged particle detections \cite{PhysRevC.87.034614,Wallace2007,must}.
The properties of the GAGG(Ce) scintillator are compared with those of the 
NaI(Tl) and CsI(Tl) scintillators in Table \ref{tab_scinti} \cite{Tamagawa2015}.

The NaI(Tl) scintillator has a light output of about
40,000 photons/MeV.
The scintillation light of NaI(Tl) can be efficiently detected with
a photomultiplier tube owing to its emission peak located at 415 nm.
The decay time of the NaI(Tl) scintillator is faster than that of the
CsI(Tl) scintillator, making it tolerant of high-counting rates.
However, the NaI(Tl) scintillator has a strong hygroscopic nature, and it
must be packed in an airtight container, which makes a
dead layer in charged particle detections.

The CsI(Tl) scintillator has a high light output
larger than the NaI(Tl) scintillator.
Because its emission wavelength matches to the Si-based photon
detectors such as a PIN photo diode and avalanche photo diode (APD),
the CsI(Tl) scintillators are often used with them.
It has a slight hygroscopic nature but can be used without a package
to prevent hygroscopy,
which makes it useful for charged particle detection.
The major disadvantage of the CsI(Tl) scintillator is its slow decay time
of the scintillation, thus it is not suitable for a measurement at
high-counting rates.

Among the three scintillators,
presented in Table \ref{tab_scinti}, the GAGG(Ce)
scintillator has the fastest decay time.
The light output and energy resolution for gamma rays are comparable with
those of the NaI(Tl) and CsI(Tl) scintillators.
Moreover, it does not have a hygroscopic nature and can be used without
any package.
Therefore, the GAGG(Ce) scintillator would be useful
for a charged particle detection at high-counting rates.
However, the response of the GAGG(Ce) scintillator to charged
particles has not been fully examined.
The particle identification between
alpha particles
at $E_{\alpha}=5.48$ MeV from an $^{241}$Am source and gamma rays
at $E_{\gamma}=662$ keV from a $^{137}$Cs source
by a pulse shape discrimination
was investigated in Refs. \cite{Kobayashi2012,Tamagawa2015}.
The light output and non-proportionality for low-energy alpha particles
at 1.5--8.8 MeV were reported in Ref. \cite{Sibczynski2018}.

\begin{table}[htb]
  \caption{Properties of the GAGG(Ce), NaI(Tl), and CsI(Tl) scintillators
    \cite{Tamagawa2015}.}
  \label{tab_scinti}
  \centering
  \begin{tabular}{|c|c|c|c|}  \hline
    & GAGG(Ce) & NaI(Tl) & CsI(Tl)\\ \hline
    Density (g/cm$^{3}$) & 6.63 & 3.67 & 4.53\\
    Light output (photons/MeV) & 46,000 & 40,000 & 50,000\\
    Emission-peak wavelength (nm) & 530 & 415 & 540\\
    Decay time (ns) & 95(79$\%$), 351(21$\%$) & 230 & 680\\
    Hygroscopic nature & None & Strong & Slight\\
    \hline
  \end{tabular}
\end{table}

In this article, we report the responses
of the GAGG(Ce) scintillator to
protons and alpha particles at $E_{p}=5$--68 MeV and $E_{\alpha}=8$--54 MeV.
The results were compared with those of the CsI(Tl) scintillator.
First, we acquired the pulse shapes from the GAGG(Ce) and CsI(Tl) scintillators 
for gamma rays. 
Then we investigated the pulse heights and the energy resolutions
with changing the time constants of a shaping amplifier.
Finally, we obtained the incident-energy dependence of the light outputs 
using the protons and alpha particles.
The scintillation efficiencies $dL/dE$ of the GAGG(Ce) and CsI(Tl)
scintillators were analyzed as a function of the linear energy transfer (LET) $dE/dx$.

\section{Performance test with a $^{137}$Cs gamma-ray source}
\label{sec:gamma}
\begin{figure}[htb]
  \centering
  \includegraphics[width=50mm]{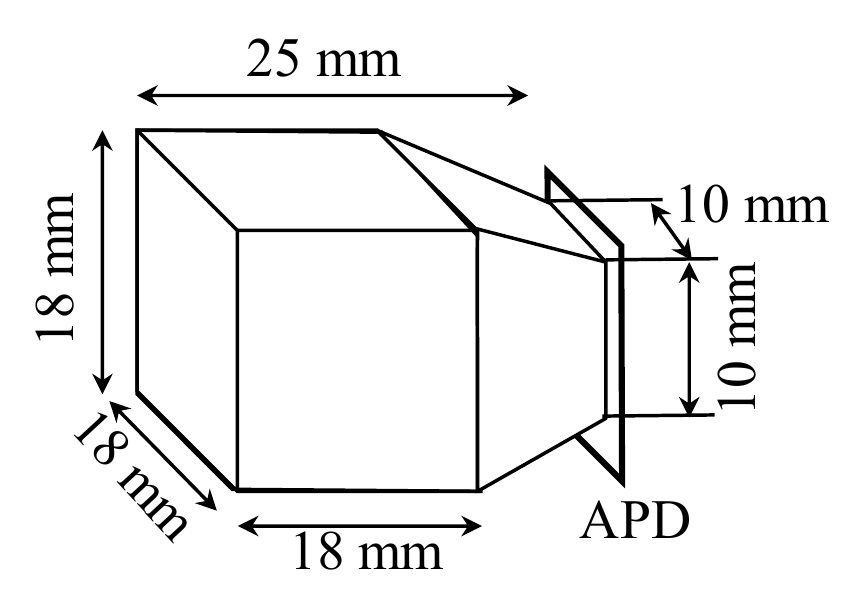}
  \caption{Geometry of the GAGG(Ce) and CsI(Tl) scintillators.}
  \label{fig_gagg_geom}
\end{figure}

The responses of the GAGG(Ce) and CsI(Tl) scintillators
to gamma rays were investigated using a $^{137}$Cs gamma-ray source.
Figure \ref{fig_gagg_geom} shows the geometry of the GAGG(Ce) and CsI(Tl)
scintillators used in the measurement.
The crystals have sizes of 18 $\times$ 18 $\times$ 25 mm$^{3}$.
The back side of the crystals was tapered so that
the crystals can be attached to
photon detectors with a sensitive area of 10 $\times$ 10 mm$^{2}$.
The photon detectors were HAMAMATSU S8664-1010 APDs.
These crystals were frosted on the surface and
wrapped with 65-$\mu$m-thick enhanced specular reflector (ESR)
films \cite{esr} to increase the light collection efficiency.

The electrical signal from the APD was amplified by a Mesytec MPR-16 preamplifier followed
by a Mesytec MSCF-16 shaping amplifier.
The decay time of the preamplifier is 25 $\mu$s.
The shaping amplifier has a 5th order CR-RC$^{5}$ filter circuit with a selectable
time constant of 0.25, 0.50, 1.0, and 2.0 $\mu$s.
The MSCF-16 shaping amplifier also has a function of a fast amplifier
with a shorter time constant for timing measurements.
The pulse height from the MSCF-16 shaper output was measured with a Mesytec
MADC-32 peak sensing analog to digital converter (ADC).

\subsection{Pulse shape}

The pulse shapes from the preamplifier and shaping amplifier
for 662-keV gamma rays were acquired with an oscilloscope.
The time constant of the shaping amplifier was set at 250 ns.
The differentiation and integration time constants of the fast amplifier
were set at $\tau_{\mathrm{dif}}=70$ ns
and $\tau_{\mathrm{int}}=20$ ns, respectively.

\begin{figure}[htb]
  \centering
  \includegraphics[width=80mm]{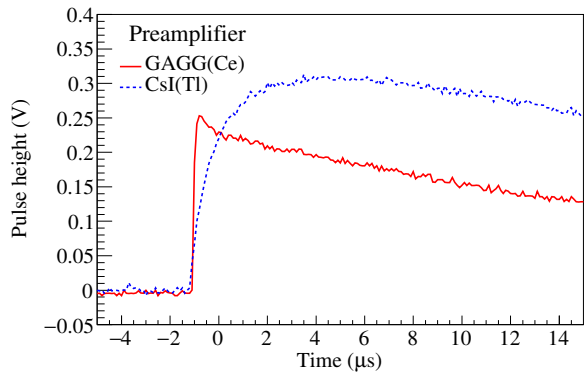}
  \caption{Pulse shapes of the preamplifier outputs for 662-keV gamma rays
    acquired with an oscilloscope.
  The red solid and blue dashed lines represent the GAGG(Ce) and CsI(Tl)
  scintillators, respectively.}
  \label{fig_preamp}
\end{figure}

The output signals from the preamplifier are shown in Fig. \ref{fig_preamp}.
The red solid and blue dashed lines represent the pulse shapes of the
GAGG(Ce) and CsI(Tl) scintillators, respectively.
The pulse heights of the preamplifier outputs in Fig. \ref{fig_preamp}
are comparable between the GAGG(Ce) and CsI(Tl) scintillators.
This means that the light outputs from the GAGG(Ce) and CsI(Tl) scintillators
for gamma rays are comparable.
We define the signal rise time as the period of time during which
the pulse height changes from 10$\%$ to 90$\%$ of the peak.
The rise time of the GAGG(Ce) scintillator was 0.13 $\mu$s, which is about 1/10 of that
of the CsI(Tl) scintillator of 1.86 $\mu$s.
This rise time reflects the decay times of the scintillation process.

\begin{figure}[htb]
  \begin{tabular}{lcl}
    \begin{minipage}{0.45\hsize}
      \centering
      \includegraphics[width=75mm]{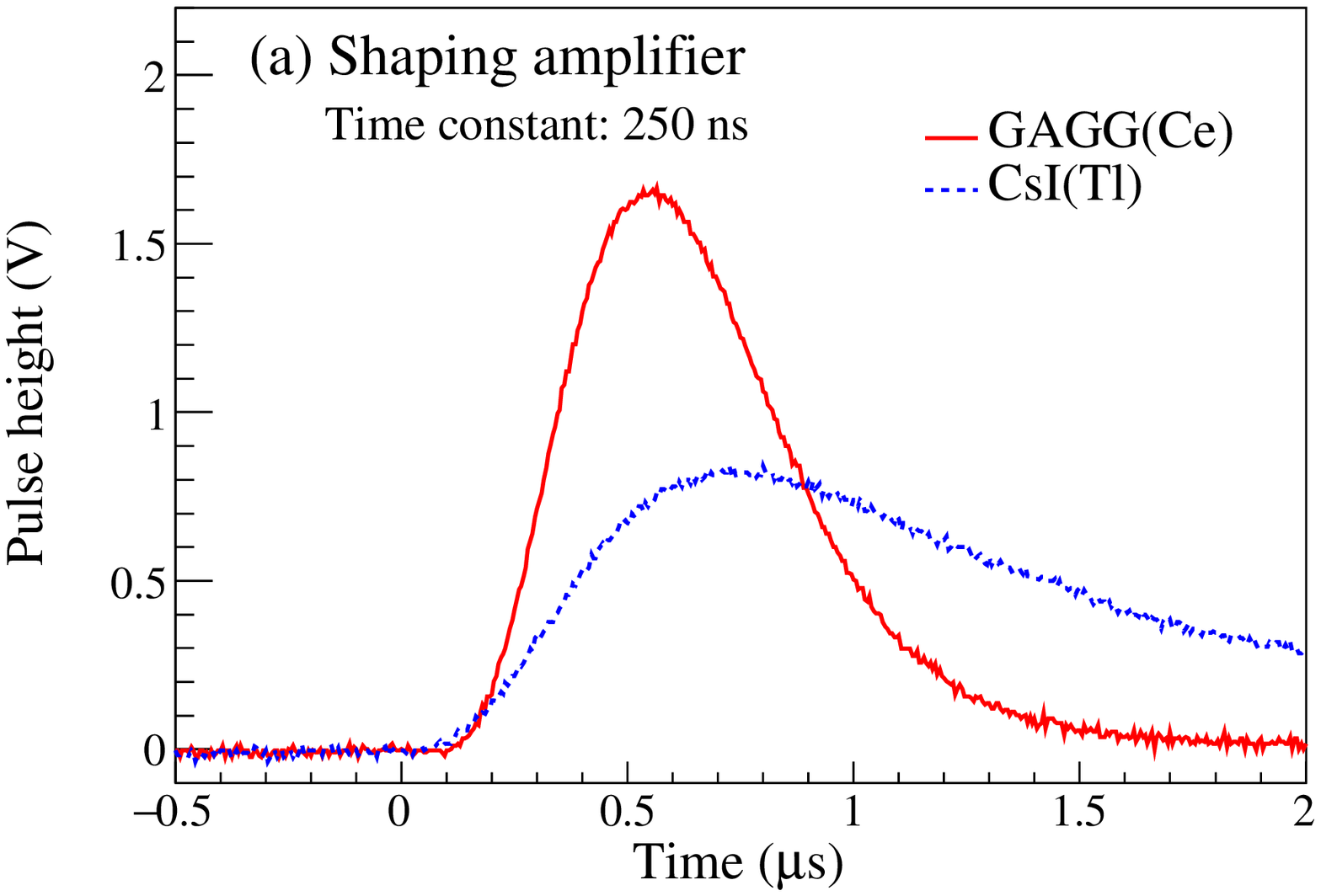}
    \end{minipage}
    & 
    \begin{minipage}{0.10\hsize}
    \end{minipage}
    &
    \begin{minipage}{0.45\hsize}
      \centering
      \includegraphics[width=75mm]{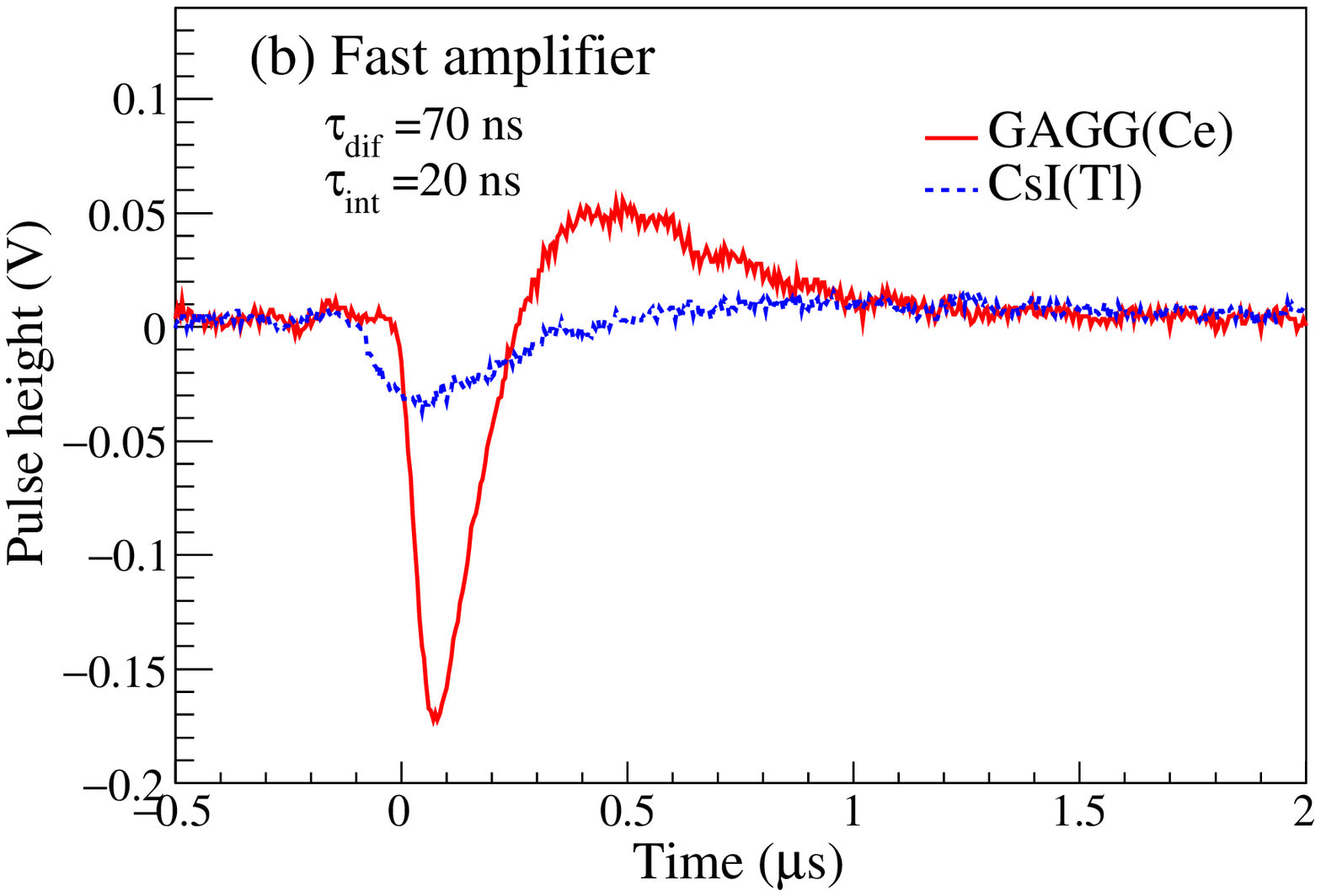}
    \end{minipage}
  \end{tabular}
  \caption{
    Comparison of (a) shaping amplifier and
    (b) fast amplifier outputs between the
    GAGG(Ce) (red solid lines) and CsI(Tl) (blue dashed lines) scintillators.
  }
  \label{fig_shaper_amp}
\end{figure}

The outputs from the fast and shaping amplifiers in MSCF-16 are shown in
Fig. \ref{fig_shaper_amp}.
Although the light outputs from the GAGG(Ce) and CsI(Tl) scintillators
are comparable,
the pulse heights of the shaping and fast amplifiers of the GAGG(Ce)
scintillator are about twice and four times larger
than those of the CsI(Tl) scintillators, respectively.
Because the decay time of the GAGG(Ce) scintillator is much shorter
than that of the CsI(Tl) scintillator,
the GAGG(Ce) scintillator has the advantage that its output signal is
efficiently processed with the shaping and fast amplifiers with a
short time constant for high-counting-rate measurements.

\subsection{Pulse height and energy resolution}

\begin{figure}[htb]
  \centering
  \includegraphics[width=80mm]{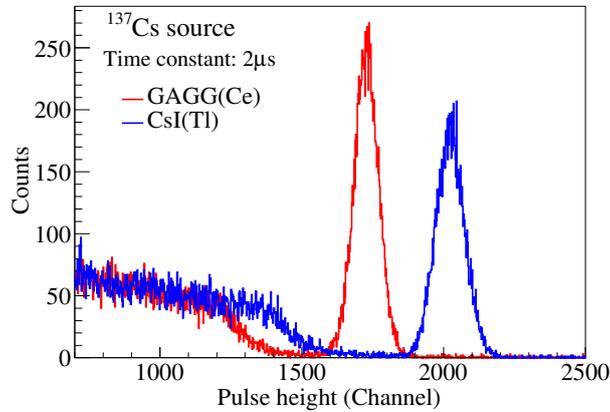}
  \caption{Pulse-height spectra of the GAGG(Ce) (red) and CsI(Tl) (blue)
    scintillators for gamma-rays from a $^{137}$Cs source.}
  \label{fig_spe}
\end{figure}

The pulse-height spectra of the scintillators
for gamma rays from the $^{137}$Cs source
are shown in Fig.~\ref{fig_spe}.
The red and blue spectra
represent the GAGG(Ce) and CsI(Tl) scintillators, respectively.
The time constant of the MSCF-16 shaping amplifier was set at 2 $\mu$s.
The pulse height and the energy resolution were estimated by fitting
the spectra by Gauss functions.

\begin{figure}[htb]
  \begin{tabular}{lcl}
    \begin{minipage}{0.45\hsize}
      \centering
      \includegraphics[width=75mm]{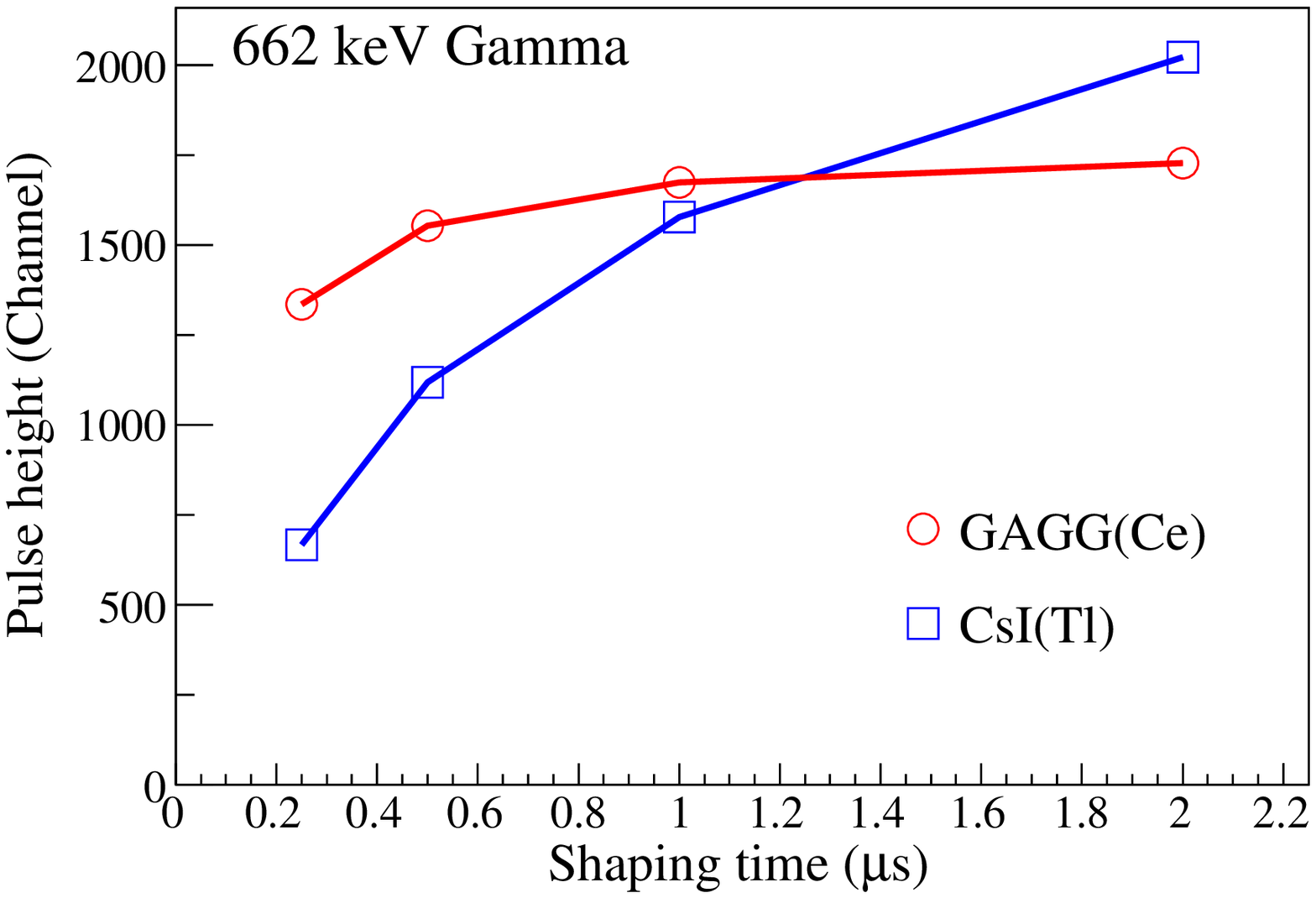}
    \end{minipage}
    & 
    \begin{minipage}{0.10\hsize}
    \end{minipage}
    &
    \begin{minipage}{0.45\hsize}
      \centering
      \includegraphics[width=75mm]{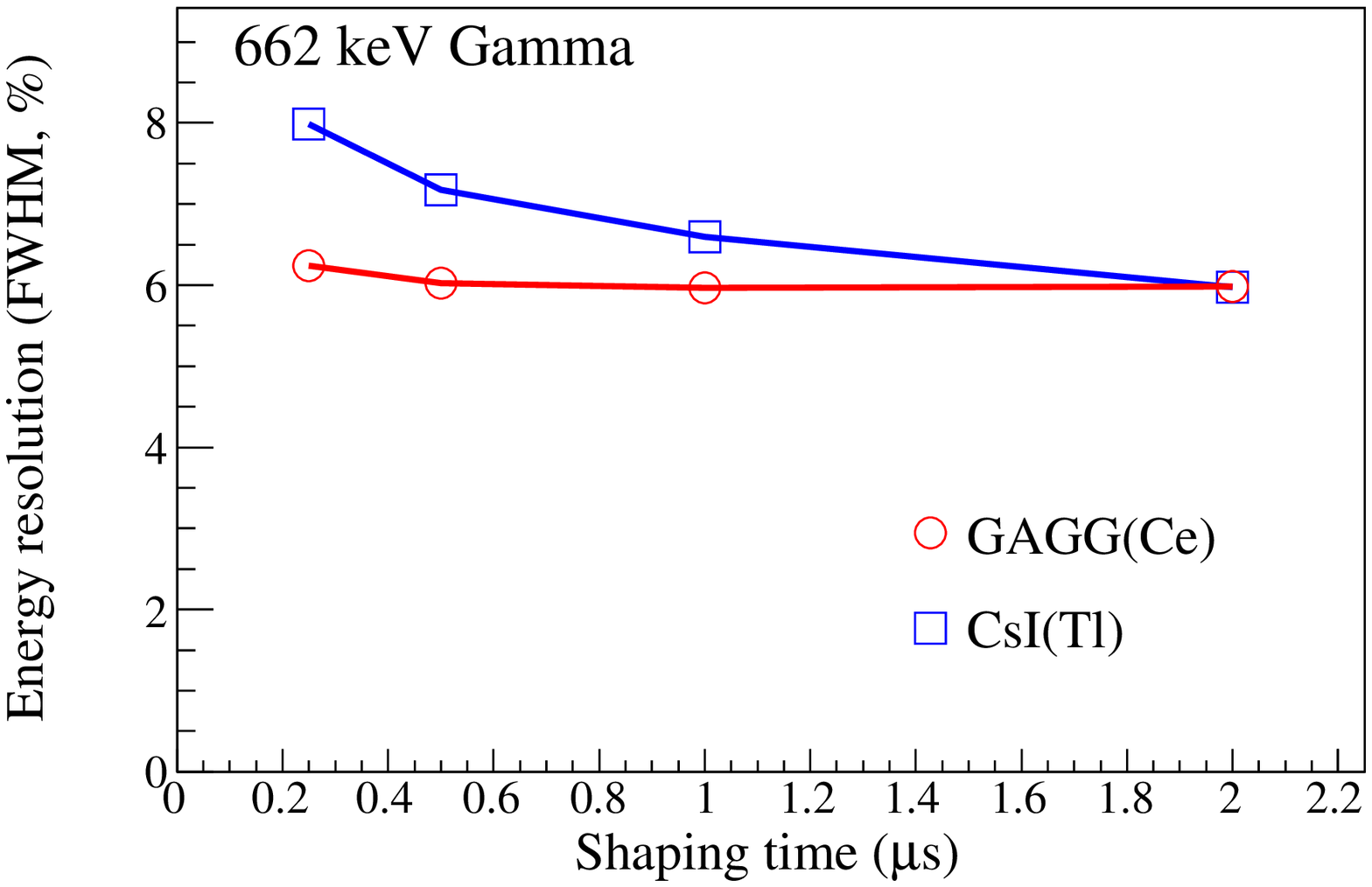}
    \end{minipage}
  \end{tabular}
  \caption{
    Pulse height
    (left) and energy resolution (right) of the scintillators
    at different shaping times of the shaping amplifier.
    The red circles and blue squares represent the GAGG(Ce) and CsI(Tl)
    scintillators, respectively.
    The solid lines connecting the data points are drawn
    for guiding eyes.
  }
  \label{fig_light_reso}
\end{figure}

Figure \ref{fig_light_reso} shows the measured
pulse heights (left) and
energy resolutions (right) in FWHM
when the time constant of the MSCF-16
shaping amplifier was set at 0.25, 0.50, 1.0, and 2.0 $\mu$s.
The red circles and the blue squares represent the GAGG(Ce) and CsI(Tl)
scintillators, respectively.
The pulse heights of both the scintillators decrease with the shorter
shaping times.
The pulse height
of the GAGG(Ce) scintillator decreases less than that of the CsI(Ce) scintillator.
The energy resolutions of the gamma-ray measurements with
the GAGG(Ce) and CsI(Tl) scintillators are comparable
when the time constant is set at 2.0 $\mu$s.
This means that the intrinsic energy resolutions of the scintillators
are almost the same.
On the other hand, when the time constants become shorter, 
the energy resolution with
the CsI(Tl) scintillator is worse than that with
the GAGG(Ce) scintillator because
the pulse height of CsI(Tl) 
from the shaping amplifier with a short time constant
is no longer sufficiently larger than
the photodetector noise.
The energy resolution with
the GAGG(Ce) scintillator shows little change with
a shorter time constant of the amplifier, making it useful in high-counting-rates
experiments where the time constant must be short to avoid pile-ups.

\section{Performance test with proton and alpha beams}
We investigated the responses of the GAGG(Ce) and CsI(Tl) scintillators to
charged particles using 70-MeV proton and alpha beams.
The beams were scattered from targets and the scattered particles were detected
by the scintillators.
The energy of the particles at the detectors was changed by placing the detectors
at different angles.
We measured the relative light outputs
of the scintillators as a function of the incident energy.

\subsection{Experimental procedure}
\begin{figure}[htb]
  \centering
  \includegraphics[width=65mm]{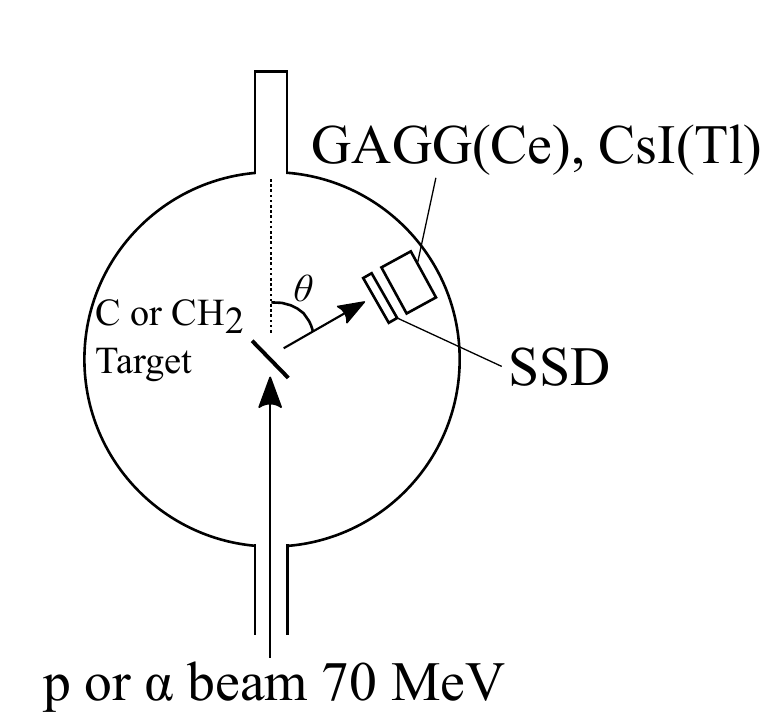}
  \caption{Experimental setup in the scattering chamber at the CYRIC 41 course.}
  \label{fig_detector}
\end{figure}

The measurement was carried out at the 41 course of
Cyclotron and Radioisotope Center (CYRIC), Tohoku University.
Figure \ref{fig_detector} shows a schematic view of the experimental setup.
The proton and alpha beams were accelerated to 70 MeV with a $K=110$ MeV
azimuthally varying-field cyclotron and
focused on a natural carbon or
a CH$_{2}$ foil target installed at the center of the scattering chamber.
The scattered particles were detected by an $E$--$\Delta E$ telescope
consisting of the GAGG(Ce) and CsI(Tl) scintillators and a silicon strip
detector (SSD).

\begin{figure}[htb]
    \begin{minipage}{0.45\hsize}
      \centering
      \includegraphics[width=75mm]{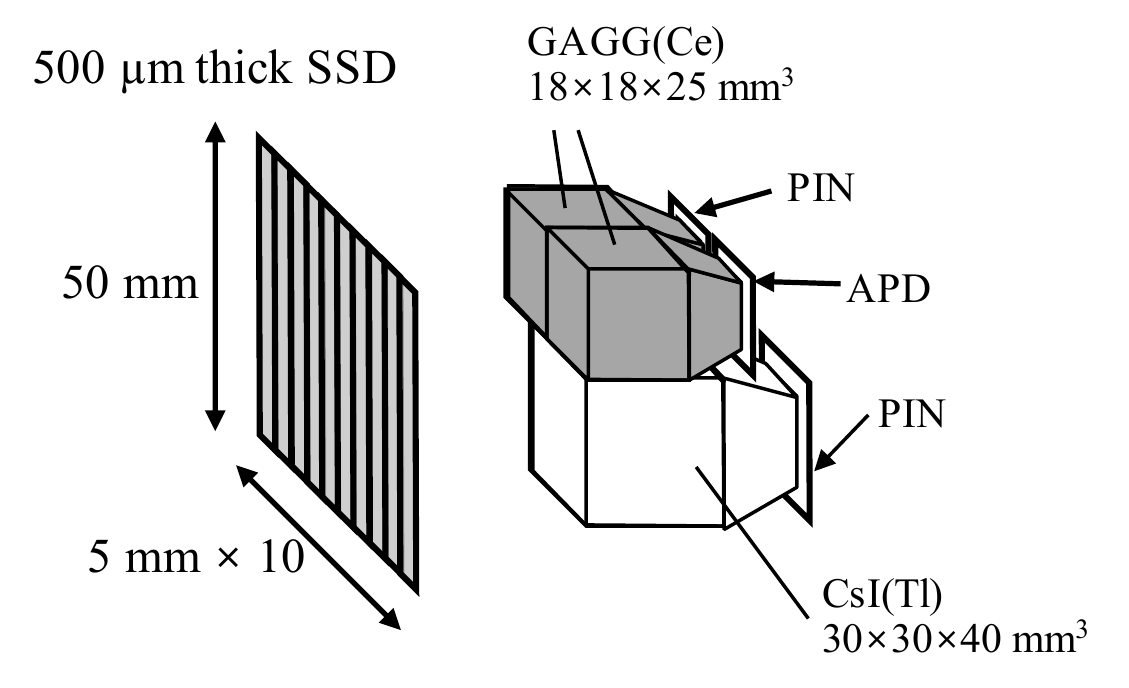}
    \end{minipage}
    \hfill
    \begin{minipage}{0.45\hsize}
      \centering
      \includegraphics[width=60mm]{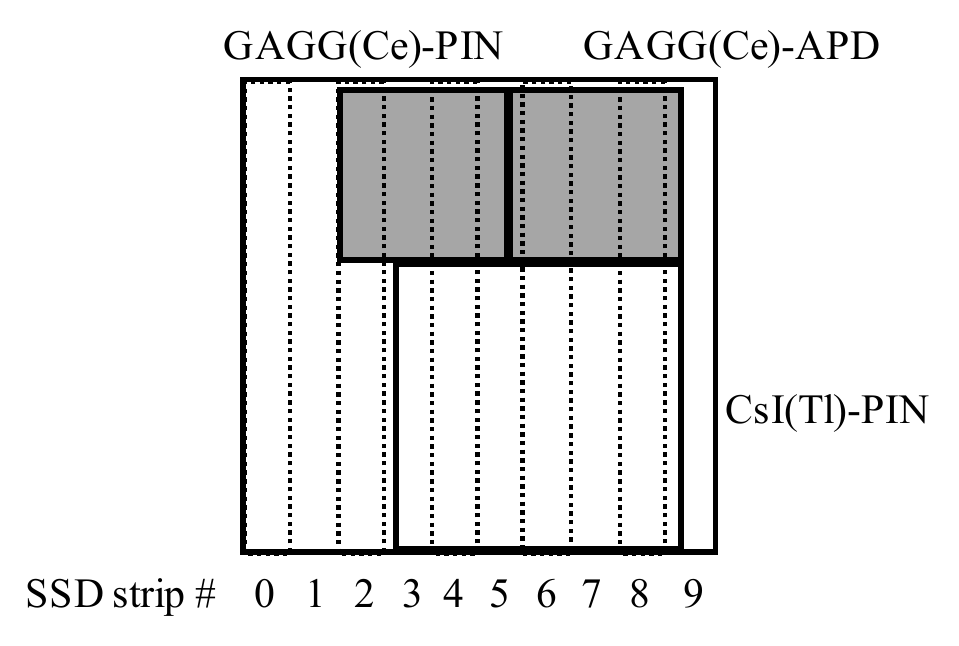}
    \end{minipage}
  \caption{
    Schematic view of the $E$--$\Delta E$ telescope.
    The left figure shows the perspective view and the right figure shows the
    front view.
  }
  \label{fig_side_front}
\end{figure}

The setup of the $E$--$\Delta E$ telescope is shown in
Fig. \ref{fig_side_front}.
For the $\Delta E$ detector at the front, we used a SSD with a
thickness of 500 $\mu$m.
The sensitive area of the SSD was 50 $\times$ 50 mm$^{2}$.
The SSD sensitive area was divided into 10 vertical strips
whose widths are 5 mm.
The SSD was used to measure the energy loss and the hit position
of the charged particles on the scintillators.
The three strips ($\#$2--$\#$4 or $\#$6--$\#$8) on the SSD were used for the
GAGG(Ce) scintillators, and the five strips ($\#$4--$\#$8) were used for
the CsI(Tl) scintillator.

For the $E$ detectors, we used two GAGG(Ce) and one CsI(Tl) scintillators.
The GAGG(Ce) scintillators were the same as those used in 
the gamma-ray measurement reported in Sec. \ref{sec:gamma}.
One GAGG(Ce) was attached to the APD as in Sec. \ref{sec:gamma}
and the other was attached to a HAMAMATSU S3590-08
PIN photo diode with a sensitive area of 10 $\times$ 10 mm$^2$.
Using the two GAGG(Ce) scintillators, we confirmed that the
photodetector resolutions with the APD and
PIN photo diode were almost same for relatively high-energy particles
inducing much larger signals than the noise.
Unlike the GAGG(Ce) scintillators, the CsI(Tl) scintillator was different
from one used in the gamma-ray measurement.
The CsI(Tl) scintillator has a volume of 30 $\times$ 30 $\times$ 40 mm$^3$
and the back side of the crystal was tapered to attach to a HAMAMATSU S3204-08
PIN photo diode with a sensitive area of 18 $\times$ 18 mm$^2$.
These crystals were frosted on the surface and wrapped with
the ESR films in the same manner as in the gamma-ray measurement.
We note that the measurements using the gamma-ray source 
reported in Sec. \ref{sec:gamma} were carried out after the present experiment
in CYRIC.
Because the CsI(Tl) scintillator with the same geometry as the
GAGG(Ce) scintillator was prepared after the experiment in CYRIC,
it was not used in the present measurement.

Because the gain of the APD changes depending on temperature,
the temperature of the detectors was monitored 
with a Pt-100 thermometer during the measurement.
The temperature was stable at $23\pm0.1$ $^\circ\mathrm{C}$.
The same preamplifier, shaping amplifier, and ADC modules
as those described in Sec. \ref{sec:gamma} were used.
The preamplifier was installed inside the scattering chamber to reduce
electrical noise.

\begin{figure}[htb]
  \centering
  \includegraphics[width=140mm]{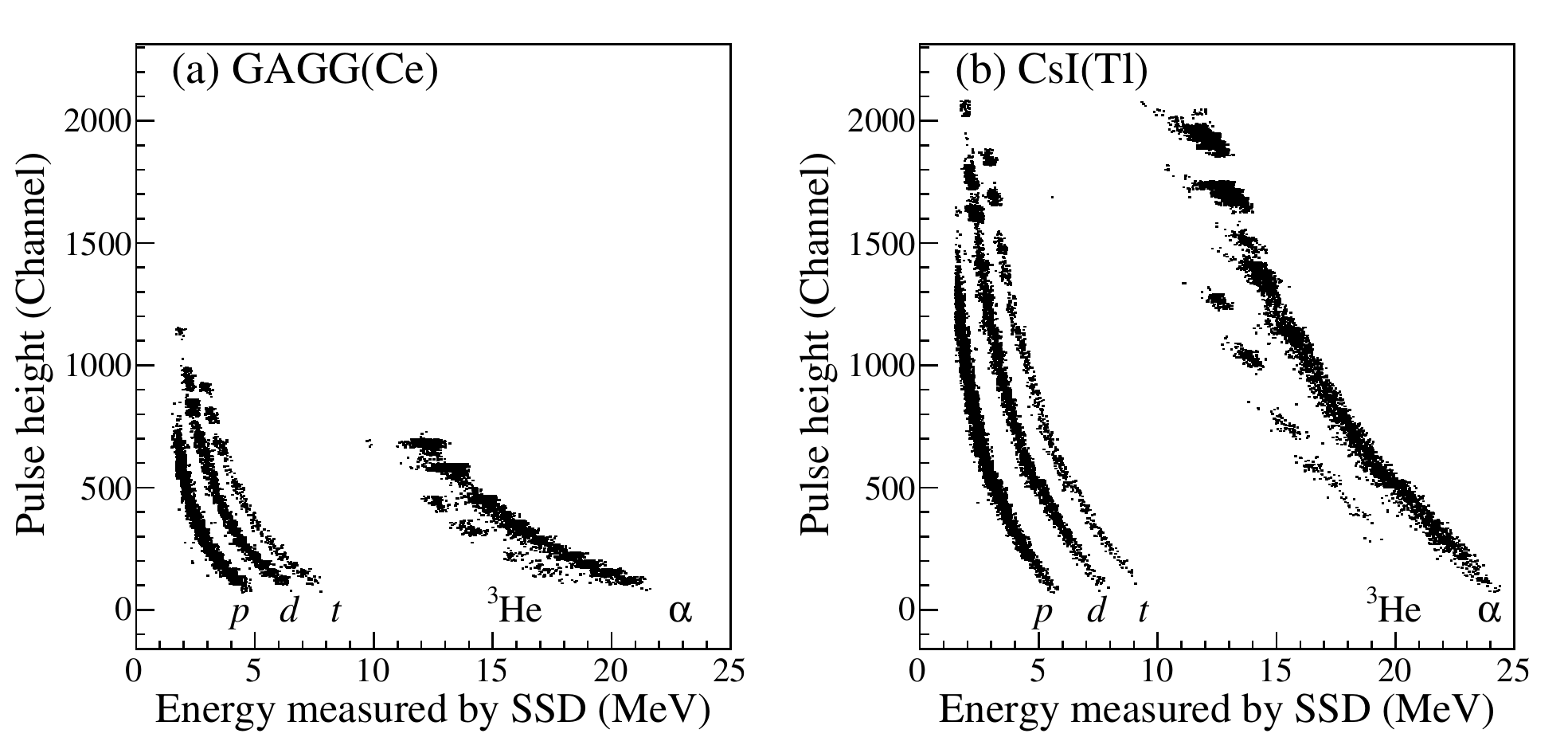}
  \caption{
    Correlation between the pulse height
    of the output signal
    from the shaping amplifier for the scintillators
    and the energy measured by the SSD.
    (a) GAGG(Ce) and (b) CsI(Tl).
  }
  \label{fig_pid}
\end{figure}

Figure \ref{fig_pid} shows the
pulse height
of the output signals from the shaping amplifier for
the GAGG(Ce) and CsI(Tl) scintillators versus the energy measured by the SSD
when the alpha beams bombarded the CH$_{2}$ target.
The left and right figures represent the GAGG(Ce) and CsI(Tl) scintillators,
respectively.
In both of the scintillators, events due to protons, deuterons,
tritons, $^{3}$He, and alpha particles were clearly separated.
In the following analysis, we selected only the proton and
alpha-particle events.

By changing the angle of the detectors ($\theta$ in Fig. \ref{fig_detector}),
the energy of the incident particles to the scintillators was
varied in an increment by about 5 MeV.
For the measurement with the proton beam, we analyzed the $p+p$ elastic scattering
and the $p+ ^{12}$C elastic scattering.
For the measurement with the alpha beam, we analyzed the 
$\alpha+ ^{12}$C elastic scattering and 
the $\alpha+ ^{12}$C inelastic scattering to the $2_{1}^{+}$ state 
at $E_{x}=4.44$ MeV.
The incident energies to the scintillators were calculated from the scattering
kinematics taking into account the energy losses through the SSD and the ESR film.
The energy losses were estimated using
the SRIM simulation code \cite{srim}.

\subsection{Pulse height and energy resolution}
\begin{figure}[htb]
  \begin{tabular}{lcl}
    \begin{minipage}{0.45\hsize}
      \centering
      \includegraphics[width=75mm]{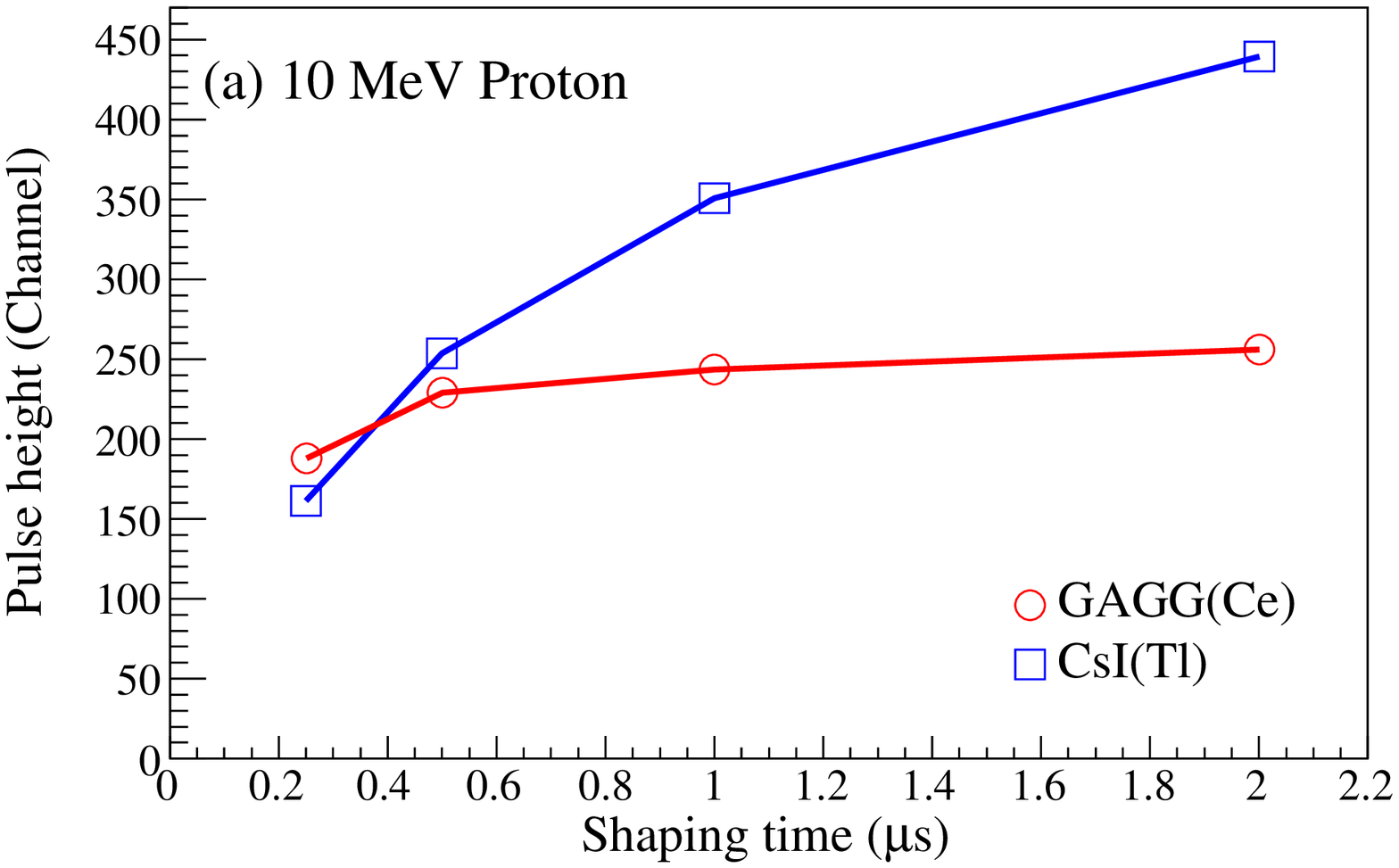}
    \end{minipage}
    & 
    \begin{minipage}{0.10\hsize}
    \end{minipage}
    &
    \begin{minipage}{0.45\hsize}
      \centering
      \includegraphics[width=75mm]{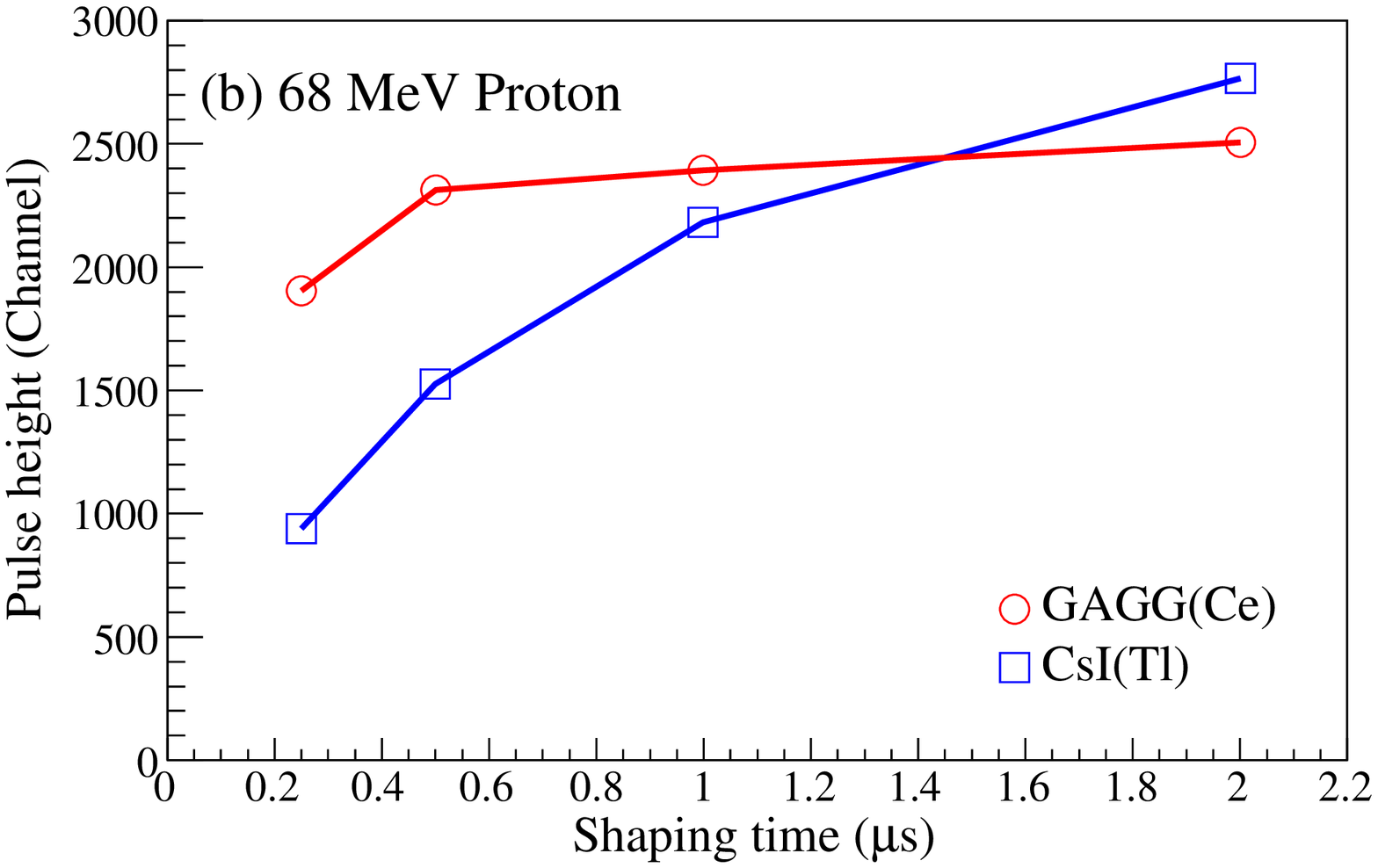}
    \end{minipage}
  \end{tabular}
  \caption{
    Pulse heights of the output signals from the shaping
    amplifier
    as a function of the shaping time for 
    (a) protons at 10 MeV and (b) protons at 68 MeV.
    The red circles represent the GAGG(Ce) scintillator and the blue squares
    represent the CsI(Tl) scintillator.
    The solid lines connecting the data points are drawn
    for guiding eyes.
  }
  \label{fig_proton_light}
\end{figure}

\begin{figure}[htb]
  \begin{tabular}{lcl}
    \begin{minipage}{0.45\hsize}
      \centering
      \includegraphics[width=75mm]{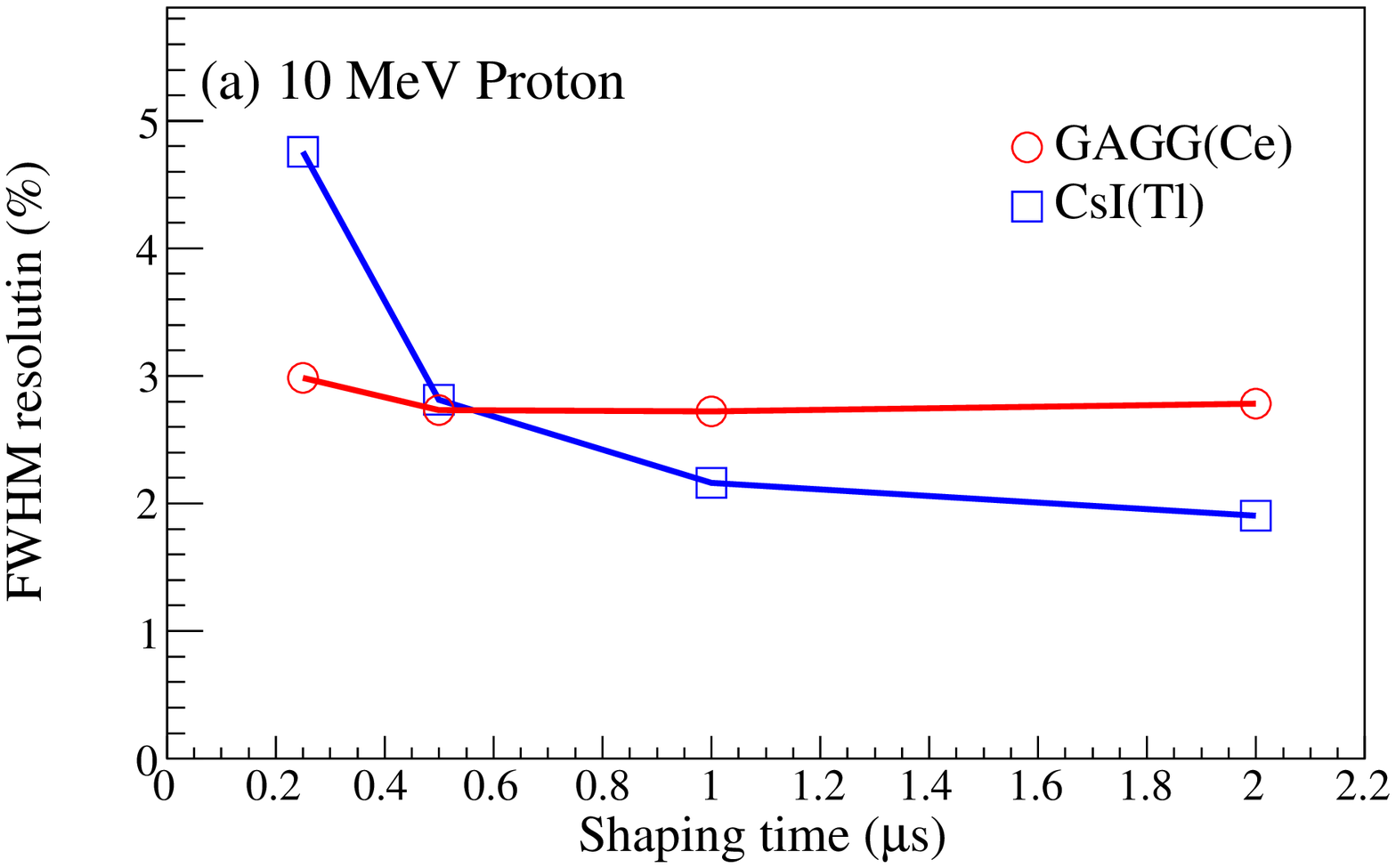}
    \end{minipage}
    & 
    \begin{minipage}{0.10\hsize}
    \end{minipage}
    &
    \begin{minipage}{0.45\hsize}
      \centering
      \includegraphics[width=75mm]{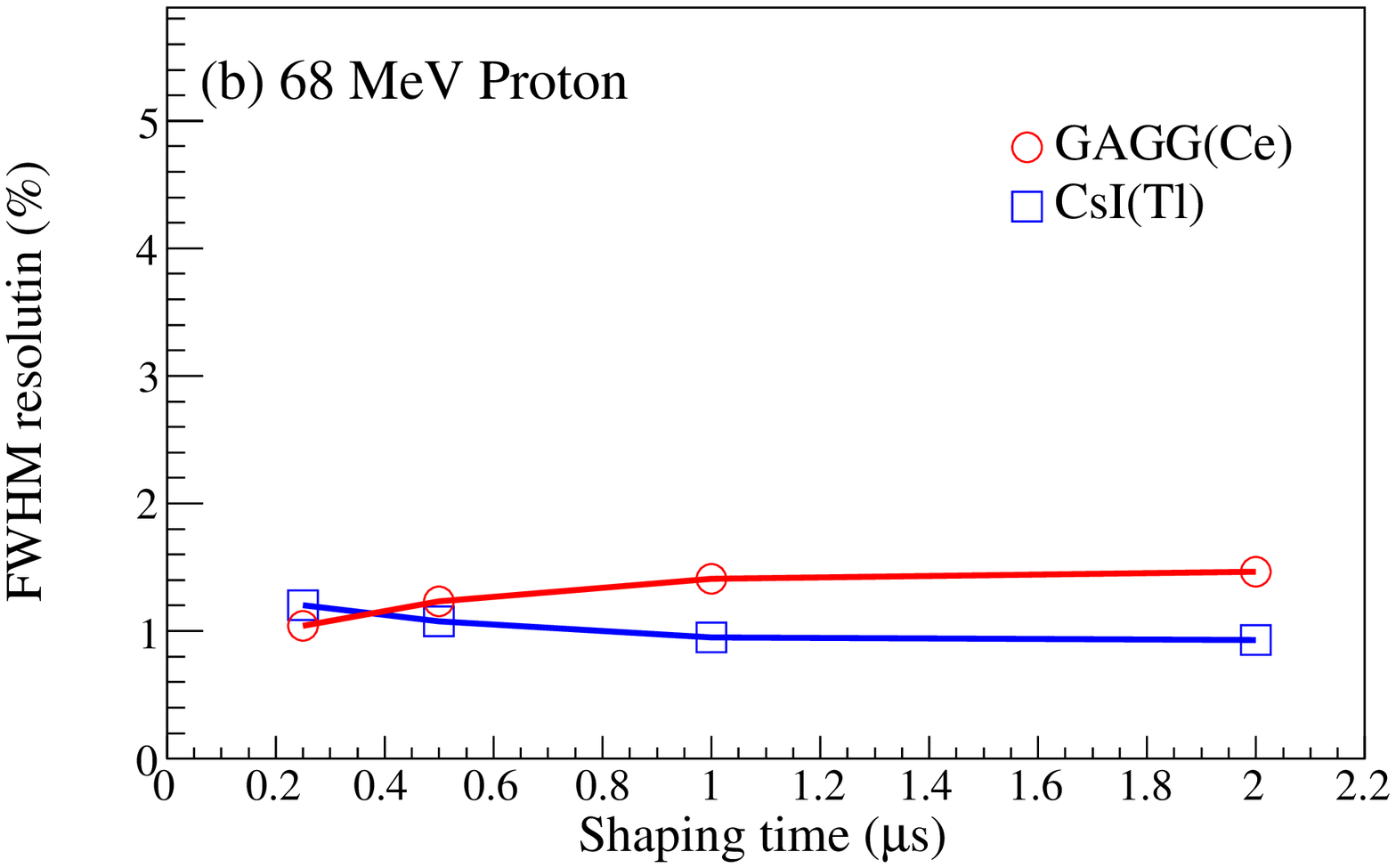}
    \end{minipage}
  \end{tabular}
  \caption{
    Energy resolution of the GAGG(Ce) and CsI(Tl) scintillators at FWHM
    for (a) protons at 10 MeV and (b) protons at 68 MeV.
    The red circles represent the GAGG(Ce) scintillator and the
    blue squares represent the CsI(Tl) scintillator.
    The solid lines connecting the data points are drawn
    for guiding eyes.
  }
  \label{fig_proton_reso}
\end{figure}

Figures \ref{fig_proton_light} and \ref{fig_proton_reso}
present the shaping time dependencies of the pulse heights
and the energy resolutions (FWHM) for protons at
(a) 10 MeV and (b) 68 MeV.
The red circles and the blue squares represent the GAGG(Ce)
and CsI(Tl) scintillators, respectively.
The shaping time was set at 0.25, 0.50, 1.0, and 2.0 $\mu$s.

The pulse height of the CsI(Tl) scintillator drastically decreases
with the shorter shaping times compared with the GAGG(Ce) scintillator.
The significant reduction of the pulse height of the CsI(Tl)
scintillator is due to its decay time longer than the GAGG(Ce) scintillator.
This reduction was observed both for protons and gamma-rays
as seen in Figs. \ref{fig_light_reso} and \ref{fig_proton_light}.
The slopes of the CsI(Tl) scintillator for the
protons at 68 MeV and gamma rays are slightly steeper than that for the protons
at 10 MeV.
This is explained by the fact that the decay time of the CsI(Tl) scintillator
becomes slightly longer as the LET decreases \cite{Alderighi2002}.

When the shaping time is set at 2 $\mu$s,
the pulse height of the GAGG(Ce) scintillator
for the 68-MeV protons 
is slightly smaller than that of the CsI(Tl) scintillator
as seen in Fig.~\ref{fig_proton_light}~(b).
On the other hand, the pulse height of the GAGG(Ce) scintillator
for the 10-MeV protons 
is much smaller compared with the CsI(Tl) scintillator
as seen in Fig.~\ref{fig_proton_light}~(a).
This suggests that the light output of the GAGG(Ce) scintillator quenches as
the LET becomes large.
This quenching effect is discussed in detail in Sec.~\ref{sec_quench}.

When the pulse height is sufficiently high, 
the energy resolution is determined by the statistical fluctuation of the 
scintillation process rather than the
photodetector noise.
This is why the energy resolutions of the GAGG(Ce) and CsI(Tl) scintillators 
for the 68-MeV protons 
are less dependent on the shaping times
as seen in Fig.~\ref{fig_proton_reso}~(b).
When the proton energy is at 10 MeV, however, 
the energy resolution of the CsI(Tl) scintillator
becomes worse with shorter
shaping time as seen in Fig.~\ref{fig_proton_reso}~(a)
affected by the noise
because the pulse height of the CsI(Tl) scintillator severely decreases.
The energy resolution of the GAGG(Ce) scintillator
at 68 MeV becomes slightly better with shorter shaping time
because the dominant frequency component of the noise
is filtered out at the shorter shaping time.

\subsection{Relative light output for protons and alpha particles}
\begin{figure}[htb]
  \centering
  \includegraphics[width=80mm]{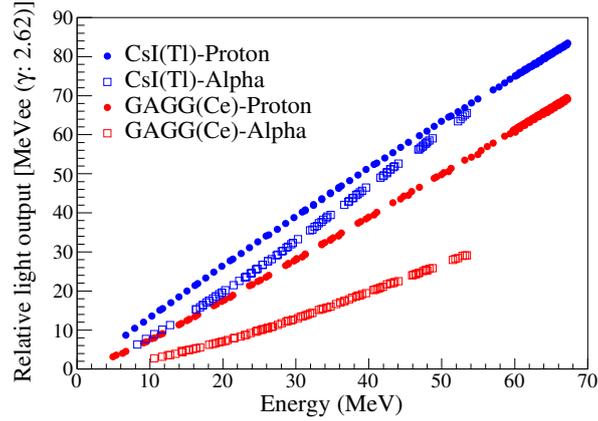}
  \caption{
    Relative light outputs for protons and alpha particles
    from the GAGG(Ce) and the CsI(Tl)
    scintillators in units of MeVee as a function of the incident energy.
    The red solid circles and open squares represent the
    light outputs from the GAGG(Ce) scintillator for protons
    and alpha particles, respectively.
    The blue solid circles and open squares represent the
    light outputs from the CsI(Tl) scintillator for protons and alpha particles.
    The relative light outputs were calibrated by a gamma ray at 2.62 MeV.
  }
  \label{fig_mevee}
\end{figure}

The light outputs from the GAGG(Ce) and CsI(Tl) scintillators
for protons and alpha particles can be
evaluated as relative values to those
for electrons in units of MeV electron equivalent (MeVee).
1~MeVee corresponds to the light output when 1 MeV of energy is given
to a scintillator by electrons.
We determined the relative light outputs
for protons and alpha particles by comparing the pulse heights with
that for a gamma ray at 2.62 MeV emitted 
after the beta decay of $^{208}$Tl from
the thorium series under the assumption that the light
output for the gamma ray is 2.62 MeVee.
The relative light outputs as a function of the incident energy
for the protons at $E_p=5$--68 MeV and alpha particles at
$E_{\alpha}=8$--54 MeV 
are shown in Fig. \ref{fig_mevee}.
The red solid circles and open squares
represent the relative light outputs from the GAGG(Ce) scintillator 
for protons and alpha particles, respectively,
whereas the blue solid circles and open squares
represent those from the CsI(Tl) scintillator for the protons and alpha particles.
The unit of the vertical axis is given in MeVee ($\gamma$: 2.62).
The suffix of ($\gamma$:~2.62) means that the calibration reference of
the relative light output is a gamma ray at 2.62 MeV.

As discussed in Ref. \cite{Tretyak2010},
the light outputs for the charged particles relative to gamma rays would be
different depending on the time constant of the shaping amplifier used.
In the present measurement, the time constants of the shaping amplifier
or the GAGG(Ce) and CsI(Tl) scintillators were 0.25 and 2.0 $\mu$s, respectively.

The relative light outputs from the GAGG(Ce) scintillator
are systematically smaller than those from the CsI(Tl) scintillator.
The relative light outputs for the alpha particles,
which provide larger LETs than protons,
are smaller than those for the protons from the same scintillator.
This shows that
the light outputs for both of the scintillators are quenched with larger LETs.

\subsection{Scintillation efficiency}\label{sec_quench}
In order to evaluate the quenching effect of the scintillators,
we investigated the relation between the light output per unit
energy loss $dL/dE$ (scintillation efficiency)
and the LET $dE/dx$.
The $dL/dE$ values were obtained using the light outputs for two
different energies measured at different angles as 
\begin{equation}
  \frac{dL}{dE} = \frac{L(\theta) - L(\theta ')}
       {E(\theta) - E(\theta ')}.
\end{equation}
$L(\theta)$ and $E(\theta)$ are the relative
  light output and the incident energy to the scintillators when the telescope detector
was placed at the angle $\theta$.
The light-collection efficiency slightly changes depending on the hit
position of the incident particle on the scintillator.
Therefore, $L(\theta)$ and $L(\theta ')$ were
determined for particles
which hit the same strip of the SSD at different angles.

The $dE/dx$ values were evaluated at the mean energy 
$[E(\theta) + E(\theta ')]/2$ using the two different codes;
SRIM and PSTAR \cite{pstar}.
The $dE/dx$ values in the CsI(Tl) scintillator given by the SRIM code
were systematically 3$\%$ smaller than those by the PSTAR code.
In the present analysis, we used the $dE/dx$ values given by the SRIM code
because the GAGG(Ce) scintillator was not included in the material listing
of the PSTAR code.

\begin{figure}[htb]
  \centering
  \includegraphics[width=80mm]{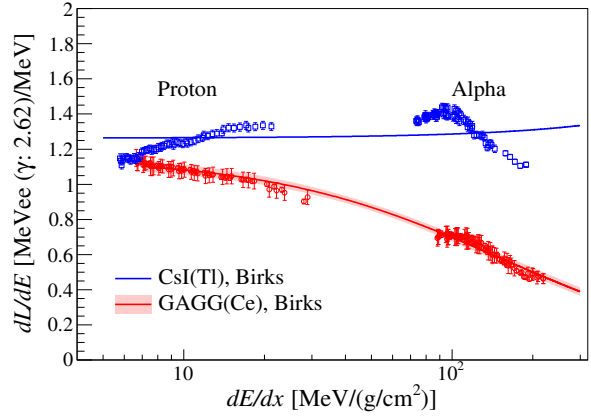}
  \caption{
    $dL/dE$ as a function of $dE/dx$.
    The red circles and the blue squares represent the GAGG(Ce) and
    CsI(Tl) scintillators, respectively.
    The solid lines represent $dL/dE$ given by the Birks formula.
    The error band around the fitted line for the GAGG(Ce) scintillator indicates
    the confidence interval of the fit at 68$\%$.
  }
  \label{fig_dlde_oribirks}
\end{figure}

The correlation between the $dL/dE$ and $dE/dx$ values is shown in
Fig. \ref{fig_dlde_oribirks}.
The red circles and the blue squares represent the GAGG(Ce) and
CsI(Tl) scintillators, respectively.
The data points at $dE/dx>30$ MeV/(g/cm$^{2}$) were obtained with the
alpha beam, while those at $dE/dx<30$ MeV/(g/cm$^{2}$) were obtained
with the proton beam.
The uncertainties of $dL/dE$ were estimated from the residuals  
when the data points in Fig. \ref{fig_mevee} were fitted by polynomials.
These uncertainties are considered to stem from inaccuracies of scattering 
angles caused by misalignment of the detector
position and non-uniformity of the light-collection efficiency.
It should be noted that the $dL/dE$ values for protons and alpha particles
behave as a smooth function of $dE/dx$ for each scintillator
as shown in Fig. \ref{fig_dlde_oribirks}.
The scintillation efficiency of the CsI(Tl) scintillator
is larger than unity.
Similar result is reported for $^{12}$C ions
in Ref. \cite{Koba2011}.
The scintillation efficiency of the GAGG(Ce) scintillator
is also larger than unity at $dE/dx < 20$ MeV/(g/cm$^{2}$).

It would be useful to determine an empirical formula of $dL/dE$
as a function of $dE/dx$.
Birks introduced the following formula to investigate the
scintillation efficiency of organic scintillators~\cite{Birks1951}:
\begin{equation}
  \frac{dL}{dx} = \frac{S(dE/dx)}{1 + kB(dE/dx)}.
  \label{eq_birks0}
\end{equation}
In this formula, it is assumed that the number of
scintillation photons
per unit length produced by the incident particle is proportional to the
energy loss per unit length $dE/dx$.
The denominator was empirically introduced so that the
scintillation efficiency decreases with larger $dE/dx$ values.
The factor $kB$ is called Birks factor.
This Birks formula reasonably fits the experimental results of the
anthracene scintillators.
Dividing this equation by $dE/dx$, the relation between
$dL/dE$ and $dE/dx$ is obtained as
\begin{equation}
  \frac{dL}{dE} = \frac{a_{0}}{1 + a_{1}(dE/dx)}.
  \label{eq_birks1}
\end{equation}
  The Birks function was fitted to the experimental data
  of the GAGG(Ce) and CsI(Tl) scintillators
  as plotted with the solid lines in Fig. \ref{fig_dlde_oribirks}.
  The obtained parameters are tabulated in Table \ref{tab_para}.
  The Birks formula reasonably reproduces the $dL/dE$ vs $dE/dx$ plot
  of the GAGG(Ce) scintillator.
  The error band around the red solid line for the GAGG(Ce) scintillator
  indicate the confidence
  interval of the fit at 68$\%$.
  On the other hand, the Birks formula
  cannot reproduce the non-monotonous trend for the CsI(Tl) scintillator at all.

\begin{table}[htb]
  \caption{Fit parameters of the Birks, modified Birks, and Romero formulae
  for the GAGG(Ce) and CsI(Tl) scintillators.}
  \label{tab_para}
  \centering
  \begin{tabular}{|c|c|c|}  \hline 
    & GAGG(Ce) & CsI(Tl) \\ \hline
    Birks & & \\
    $a_0$ & $1.15$ & $1.27$ \\
    $a_1$ & $6.5 \times 10^{-3}$ & $-1.1 \times 10^{-4}$ \\ \hline
    Modified Birks &  & \\ 
    $a_0$ & 1.11 & 1.74 \\
    $a_1$ & $6.1 \times 10^{-3}$ & $2.6 \times 10^{-3}$ \\
    $a_{-1}$ & $-3.2 \times 10^{-1}$ & 3.7 \\ \hline
    Romero &  & \\ 
    $a_0$ &  4.68 & 4.20  \\
    $a_1$ & -5.84 & -5.32 \\
    $a_2$ &  3.81 & 3.53  \\
    $a_3$ & -1.22 & -1.11 \\ 
    $a_4$ & $ 1.89 \times 10^{-1}$ & $ 1.72 \times 10^{-1}$ \\ 
    $a_5$ & $-1.14 \times 10^{-2}$ & $-1.06 \times 10^{-2}$ \\ \hline
  \end{tabular}
\end{table}

\begin{figure}[htb]
  \centering
  \includegraphics[width=80mm]{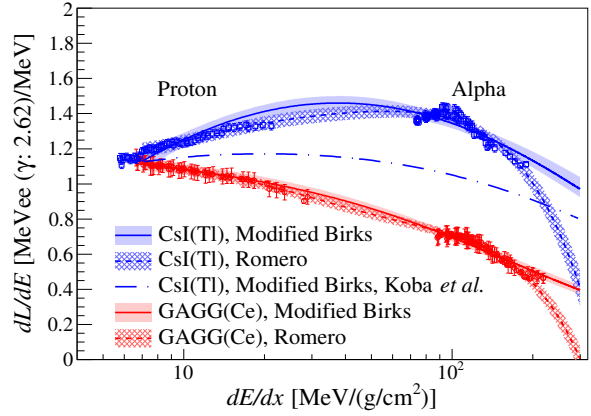}
  \caption{
    Same as Fig. \ref{fig_dlde_oribirks}, but the experimental data was fitted by
    the modified Birks (solid lines) and Romero formulae (dashed lines).
    The blue dash-dotted line represents $dL/dE$ given by the 
    modified Birks formula for the CsI(Tl) scintillator
    reported by Koba {\it{et al.}} \cite{Koba2011}.    
  }
  \label{fig_dlde}
\end{figure}

Koba {\it{et al.}} modified the Birks formula as
\begin{equation}
  \frac{dL}{dE} = \frac{a_{0}}{1 + a_{1}(dE/dx) + a_{-1}(dE/dx)^{-1}},
  \label{eq_birks2}
\end{equation}
to reproduce the scintillation efficiency of the CsI(Tl)
scintillator \cite{Koba2011}.
Romero {\it{et al.}} introduced another empirical formula to fit
the scintillation efficiency of the NaI(Tl) scintillator \cite{Romero1991}:
\begin{equation}
  \frac{dL}{dE} = \sum_{j=0}^{5} a_{j} 
  \biggl( \ln \frac{dE}{dx} \biggl)^{j}.
  \label{eq_romero}
\end{equation}
In the present analysis,
we fitted Eqs. (\ref{eq_birks2}) and (\ref{eq_romero}) to $dL/dE$ of the
GAGG(Ce) and CsI(Tl) scintillators at $dE/dx=5$--200~MeV/(g/cm$^2$)
as shown in Fig. \ref{fig_dlde}.
The fitted lines are drawn with the solid lines (modified Birks) and
the dashed lines (Romero) associated with the error bands.
The blue dash-dotted line represents the previous fit result
with the modified Birks formula for the CsI(Tl) scintillator reported by
Koba {\it{et al.}} \cite{Koba2011}.
The present result of the CsI(Tl) scintillator is systematically
20$\%$ larger than the previous result.
This discrepancy is probably because of the concentration
of Tl doped in the CsI crystal.
Murray and Meyer reported that the $dL/dE$ values vary up to 100$\%$ with the
Tl concentration from 0.01$\%$ to 0.2$\%$ \cite{PhysRev.122.815}.

The scintillation efficiency of the CsI(Tl) scintillator
is more stable over the wide $dE/dx$ range than the GAGG(Ce) scintillator,
making the CsI(Tl) scintillator useful as a charged particle detector.
However, its slow response would cause a pile-up at high-counting rates.
The GAGG(Ce) scintillator is a good candidate for light charged
particles such as protons and alpha particles thanks to the fast response
as well as the good energy resolution although
the scintillation efficiency notably decreases with the larger $dE/dx$.

Recently, we constructed a GAGG(Ce) based light-ion telescope (Gion)
which consists of 24 GAGG(Ce) scintillators with the same geometry as
shown in Fig. \ref{fig_gagg_geom} and a double-sided silicon strip detector.
Gion was successfully employed to detect recoil protons emitted from
$^{12}$C($p,p'$) reaction in inverse kinematics.
Using Gion, the rare radiative-decay probability of the $3_{1}^{-}$ state
at $E_{x}=9.64$~MeV in $^{12}$C was determined to estimate the triple alpha
reaction rate in high temperature environments \cite{Tsumura2021}.

\section{Summary}
The light output and energy resolution of the GAGG(Ce) and CsI(Tl)
scintillators were compared using
662-keV gamma-ray quanta from a $^{137}$Cs source, 2615-keV gamma-ray
quanta of $^{208}$Tl from environmental radioactivity, and accelerated beams.
For gamma rays,
the light outputs from the GAGG(Ce) and CsI(Tl) scintillators
are comparable.
The energy resolution
of the two scintillators are also comparable when the time constant of
the shaping amplifier is 2 $\mu$s.
However, at shorter time constants, the energy resolution of the
GAGG(Ce) scintillator is better than that of the CsI(Tl) scintillator
because the GAGG(Ce) scintillator has a significantly shorter decay
time of the scintillation than the CsI(Tl) scintillator.
These results demonstrated that the GAGG(Ce) scintillator is 
more suitable in measurements at high-counting rates
than the CsI(Tl) scintillator.

The light outputs for charged particles were measured using
protons at $E_{p}=5$--68~MeV and alpha particles at $E_{\alpha}=8$--54~MeV.
The empirical formulae to describe the scintillation efficiencies
$dL/dE$ as a function of $dE/dx$ were obtained at $dE/dx=5$--200~MeV/(g/cm$^2$).

\acknowledgments
The authors would like to acknowledge the cyclotron crews at CYRIC
for the stable operation of the cyclotron facilities.
This research was supported by JSPS KAKENHI, 
Grants No. JP14J00949, JP15H02091, JP19H02422, and JP20K22351,
and the GIMRT Program of the Institute for Materials Research, Tohoku University
(Proposal No. 15K0071 and 16K0060).

\bibliographystyle{JHEP}
\bibliography{mybibfile}

\end{document}